\documentclass{elsarticle}
\usepackage{color,lineno,hyperref,amsfonts,amsbsy,amssymb,amsmath,graphicx,inputenc}
\journal{Marine Pollution Bulletin}

\begin{document}

\begin{frontmatter}

%\title{An Interdisciplinary Approach for a Real-Time Response to a Marine Oil Spill}
%\title{An Interdisciplinary Strategy for a Real-Time Response to a Marine Oil Spill}
%\title{An   Real-Time Interdisciplinary Response to a Marine Oil Spill}
%\title{Efficient strategies  Real-Time Response to a Marine Oil Spill}
\title{A Dynamical Systems Perspective for a Real-Time Response to a Marine Oil Spill}

\author[mymainaddress]{V. J. Garc\'ia-Garrido\corref{mycorrespondingauthor}}
\cortext[mycorrespondingauthor]{Corresponding author}
\ead{victor.garcia@icmat.es}

\author[mysecondaryaddress]{A. Ramos}
\author[mymainaddress]{A. M. Mancho}
\author[mysecondaryaddress]{J. Coca}
\author[mythirdaddress]{S. Wiggins}

\address[mymainaddress]{Instituto de Ciencias Matem\'aticas, CSIC-UAM-UC3M-UCM, \\ C/ Nicol\'as Cabrera 15, Campus Cantoblanco UAM, 28049, Madrid, Spain.}
\address[mysecondaryaddress]{Divisi\'on de Rob\'otica y Oceanograf\'ia Computacional, IUSIANI,\\
Universidad de Las Palmas de Gran Canaria, Las Palmas de Gran Canaria, Spain.}
\address[mythirdaddress]{School of Mathematics, University of Bristol,\\
Bristol BS8 1TW, United Kingdom.}

\begin{abstract}
This paper discusses the combined use of tools from dynamical systems theory and remote sensing techniques and shows how they are effective instruments which may greatly contribute to the decision making protocols of the emergency services for the real-time management of oil spills. This work presents the successful interplay of these  techniques for a recent situation, the sinking of the Oleg Naydenov fishing ship that took place in Spain, close to the Canary Islands, in April 2015. 
\end{abstract}


\end{frontmatter}

%\linenumbers

\section{Introduction}
On Saturday 11th April 2015, after refuelling 1500 tons of IFO 380 oil in the Port of Las Palmas de Gran Canaria island (Spain), the Russian fishing trawler Oleg Naydenov caught fire at 13:30 UTC. Spanish authorities towed the ship out of the port facilities for security reasons with the tugboat Miguel de Cervantes. This manoeuvre was intended to tow the ship into southern waters until the fire was extinguished, after which it could be brought back to port securely. Unfortunately, the fishing ship eventually sank on the night of the 14th April, in waters 2700 meters deep. After the sinkage, on the 16th of April, several oil slicks were spotted on the sea surface. Following these events, actions were taken by the local and national authorities and the emergency services to diminish the environmental damage (information is available at http://www.fomento.gob.es/MFOMBPRENSA/). Three tugboats and a recovery platform carried Remote Operated Vehicles (ROVs) to seal the ship at 2.7 km depth. The ROVs sealing operations began in the summer and were completed by December 2015. The plan also included the use of two coast guard aircrafts and one helicopter to patrol the SW Gran Canaria sector and gather information regarding the location and evolution of  the spill path. The flying routes included the western archipielago waters in order to search for unreported slicks. Despite these measures, several oil patches landed onto the coast of Gran Canaria on the 23rd April. At that point, the authorities activated an emergency plan to report and clean the affected areas. These tasks continued until the end of May, when the event was considered to be under control. 

This paper aims to account for the real-time evolution of the spill both from the emergency services perspective and from an interdisciplinary approach involving remote sensing techniques and dynamical systems theory. The goal is to demonstrate that these techniques, when assembled, are effective for providing a real-time response to these types of catastrophic events. To that end, the information retrieved by the emergency services is used to validate the satellite imagery techniques and the Lagrangian approach provided by dynamical systems theory. In turn, when these techniques are validated it is found that they are capable of providing important feedback to the emergency services.
 
The emergency services acquire a precise knowledge of the evolution of spills by means of a formidable, extremely time-consuming and expensive effort, which involves in situ measurements carried out by  maritime rescue ships, air observations accomplished by exploratory flights of Search and Rescue aircraft, and verification of testimony from people  in the affected area. This approach is thus made from many local observations and requires an assimilation exercise in order to obtain a global vision of the impact of the spill. 

In the unfortunate event of a marine oil spill, it is crucial to take advantage of different scientific tools and mathematical models that will help  identify the spill and provide a description of the oceanic transport processes driving the currents in the region where this environmental catastrophe occurs. Therefore, the synergy between different approaches undoubtedly plays a key role in the development of contingency plans and strategies to assess effectively these disasters, in order to reduce their ecological impact as much as possible in future events.

Remote sensing techniques allow, in contrast to the emergency services approach, a direct observation of extended areas. Currently, the optical sensors available onboard satellites allow for the detection of oil spills (see \cite{pisano} and references therein). For instance, the Synthetic Aperture Radar (SAR) is the most useful remote sensing sensor for oil spill detection \citep{pisano}, due to its high spatial resolution and all-weather and all-day capabilities. The constraints of SAR are mostly related to their swath width and revisit time. These limitations  make SAR products unsuited  for the Oleg Naydenov oil spill. The results discussed in this work are based on the outputs (images) of the Moderate Resolution Imaging Spectroradiometer (MODIS) sensor. The suitability of MODIS for oil feature detection is discussed in \cite{bul}. These authors reported that the retrieved information depends on the illumination conditions: in the case of sunglint contamination, MODIS can locate the oil spill similar to radar detection, as a sea surface roughness anomaly; whereas in the absence of sunglint, oil spill spectral properties can be extracted. The analysis of these satellite images thus requires expertise for distinguishing false positives caused by the different settings.

MODIS data from Terra (EOS AM-1) and Aqua (EOS PM-1) satellites provide daily coverage over the study area with different solar illumination patterns. In contrast, the revisit time of SAR sensors is much less frequent (for example 12 days for Sentinel-1A satellite). Additionally, atmosphere and ocean dynamics in the surrounding areas of the sinking point  constrain the SAR capabilities. The interaction of island trade winds  produce major changes in the wind stress on the sea surface, in which zones  with different intensities alternate with  areas, in the wake of the islands,  of total calm. These changing patterns further restrict the use of SAR sensors for this particular case.
 
Lastly, mathematical techniques  also provide direct information on extensive areas, and have the advantage of being able to predict the time evolution of the spill, thus providing a powerful tool for a real-time response. These techniques are based on tracking  particle trajectories, and to this end they  require the use of velocity data sets supplied by physics-based computational models which incorporate data from observations and are solved on high-resolution meshes. Obviously, the success of  this approach depends on how well the data represents the ocean state. In this study, we have used COPERNICUS IBI data, which provides ocean circulation currents  on the whole Canary Islands ocean basin.  For this problem, COPERNICUS IBI data has turned out to be remarkable for achieving this goal.

The mathematical approach we follow in this analysis goes beyond the tracking of  pollutant trajectories, as provided by  tools such as the General NOAA Operational Modeling Environment (GNOME) (see for instance \citep{mp1,mp2,mp3} for their implementation in different oil spill events). Our analysis utilizes  fundamental ideas from dynamical
systems theory  which search for sharper insights.  In particular we exploit Poincar\'e's idea of determining the geometrical structures in the phase space (for this problem, the ocean surface)  that define regions where particle trajectories have qualitatively different behaviors. The boundaries, or barriers, between these regions are mathematically realized as objects called manifolds, also known as Lagrangian coherent structures (LCS). The task of finding these geometrical structures characterizing transport processes in the context of geophysical flows is a difficult challenge, which in the literature has been addressed in several ways. Both finite size Lyapunov exponents (FSLE) \citep{vulpiani} and finite time Lyapunov exponents (FTLE) \citep{nese,shaden,sha09} have been succesfully applied into oceanic contexts \citep{emilio,birds,olas}. Another perspective is the direct computation of manifolds  \citep{physrep,msw,mm2012}, which has also provided valuable insight into oceanic problems \citep{jpo,nlpg2}. Other approaches in this field have been the geodesic and variational theories of LCS \citep{beron-vera,fh12}, the trajectory complexity measures \citep{rypina}, mesohyperbolicity measures and ergodic partitions \citep{mezic3,mezic} and transfer operator methods \citep{gary, froy12}. Lagrangian tools have provided in the past interesting insights into oceanic problems related to oil spills and pollutant release. For instance, they have been applied to the release of organic contaminants in the coast of Florida \citep{Lekien} or to the Gulf oil spill disaster after the Deepwater Horizon oil rig explosion in 2010 \citep{mezic}.

In this work we illustrate the power of a recent dynamical system tool,  Lagrangian descriptors (LD), also known as the function $M$, to analyze the fate of Oleg Naydenov's spill and its potential environmental impact on the Canary Islands ecosystem. LD achieve the goal of revealing the above mentioned material barriers in the ocean surface (cf. \citep{prl,mm2012}). The choice of this tool versus others mentioned above, such as for instance FTLE, is based on the advantages discussed in the literature \cite{cnsns,rempel,alvaro,alvaro2}. These are related to its accuracy, ability to detect Lagrangian features of the true ocean state, computational efficiency and programming simplicity. The good performance of  the function $M$ is demonstrated by comparing its  outputs to those of FTLE in some chosen examples for our problem. Our analysis is completed with techniques derived from contour advection \citep{dritschel,physicad}.

This paper is organized as follows. Section 2 describes the data sources used in this work: satellite data, the COPERNICUS IBI product, and the methodology used to extract information from the data, {i.e} how the dynamical systems approach is applied to the velocity data sets. Section 3 presents the results and discussion. Finally Section 4 reports the conclusions.

\section{Data and Methodology}

\subsection{Satellite Imagery and Remote Sensing}

The tasks of the sea/air monitoring operatives were complemented throughout the event with an extensive analysis of satellite imagery and radar data in near real time from the 16th April to the 10th May (when the spill was finally confined to the sinking point). Since the ship sank close to the southwestern Gran Canaria island wake, where the sea surface is protected from the wind and the currents,  the usage of radar data from SAR passes is rather limited. Only a few Sentinel-1A and Landsat 7/8 images were available during the whole period, and their usage in near real time was prevented by clouds, the  different traces of the sinking point and  sunglint conditions. However, all the MODIS Aqua and Terra passes were available and proved to be a surprisingly useful satellite-derived tool to identify in near real time potentially contaminated areas in order to initialize the spill's forecast.

The remote sensing data used for our analysis are Level-1A (L1A) diurnal granule images of MODIS acquired with the Aqua and Terra satellites. These satellites covered the affected area of 26$^o$ to 29$^o$ N and 17$^o$ to 14$^o$ W from the 15th of April to the 15th of May 2015. The data were downloaded from the Ocean Color Web (http://oceancolor.gsfc.nasa.gov/) and the processing was conducted using the SeaWIFS Data Analysis System (SeaDAS) 7.2 (http://SeaDAS.gsfc.nasa.gov/).

Quasi-true color images were processed by means of the SeaDAS binary {\ttfamily l1mapgen}, activating the 250m resolution and atmospheric correction options. This processing technique produced natural color images that allowed for oil spill visualization attending to solar illumination and satellite viewing geometry patterns.

In order to derive geophysical variables, MODIS granule images were processed from raw L1A to a geophysical product through the generation of GEO (geolocation files) and L1B intermediate files. In order to achieve this, first the appropriate attitude and ephemerids data corresponding to each L1A file were downloaded using the SeaDAS script {\ttfamily modis\_atteph.py}. Then, geolocation files were produced using SeaDAS script {\ttfamily modis\_GEO.py}, and afterwards, the L1B files were generated with the script {\ttfamily modis\_L1B.py}. 

L1B and GEO products were used for the L2 processing, which leads to obtaining the reflectances employed in our analysis.  The L2 corresponding ancillary data were downloaded using {\ttfamily getanc.py}, and the L2 files were produced by means of the {\ttfamily l2gen} binary, selecting the 250 meters resolution with  the replication for lower resolution bands option. Finally, L2 granule images were projected onto a geographical Lat/Lon WGS84 projection using the graphical processing tool {\ttfamily gpt.sh} (reproject operator).

\subsection{COPERNICUS IBI}
The oceanic velocity data used to analyze the oil spill evolution throughout the event has been retrieved from the COPERNICUS Iberian-Biscay-Irish (IBI) Monitoring \& Forecasting Centre (available at http://marine.copernicus.eu/). This product is based on a NEMO (Nucleus for European Models of the Ocean) model \citep{madec} coupled to the LIM2 thermodynamic sea ice model. It also comprises both physical and biogeochemical components recorded since 2002. IBI data is driven with atmospheric forcing fields (which includes winds) provided by the Meteorologic Model MM5 developed by the European Centre for Medium-Range Weather Forecasts (ECMWF). The boundary conditions, which have a 6-hour update frequency, are obtained from the basin-scale open-ocean Forecast Ocean Assimilation Model (FOAM), which also provides initial and tidal forcing and includes updates of the fresh water river discharges \citep{bell,holt,sot08,sot15,maraldi,lorente}.

\begin{figure}[htbp]
 \centering
 \includegraphics[scale=.7]{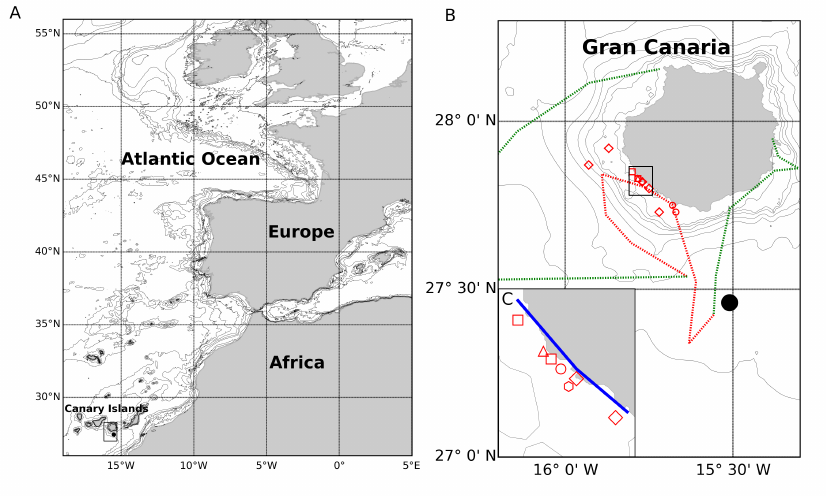}
 \caption{A) The COPERNICUS IBI domain; B and C) Study area with reported oil spill sighting points in the Gran Canaria coast and close ocean areas (triangles 23rd April 2015, squares 24th April 2015, diamonds 25th April 2015, circles 26th April 2015 and hexagons 27th April 2015). Oleg Naydenov's sinking point is showed with a filled black circle. The dashed line in B) accounts for the daily coast guard aircraft monitoring flight paths from the 23rd April to the 10th May. The green sections in the flight paths correspond to regions where no oil was reported and red accounts for oil spill reported sections. The blue line in C) along the Gran Canaria coast represents the coast guard helicopter operational track from the 23rd April to the 10th May.}
 \label{events}
\end{figure}

To address the spill evolution and help to design an effective plan of action for this emergency scenario, we use the data set IBI\_ANALYSIS\_FORECAST\_PHYS\_005\_001\_b. This data provides continuous hourly/daily estimates and 5-day forecasts of the ocean velocity fields. In particular we use the daily estimates. The horizontal resolution is of $1/36^{\circ} \; (\sim 2 {\rm km})$ on a regular Lat/Lon equirectangular projection of the Northwest African waters domain (see Fig. \ref{events}A), and daily means are available over 50 geopotential vertical levels.

Other products providing ocean currents for the area of interest are for instance the Hybrid Coordinate Ocean Model (HYCOM), or AVISO which provides mainly geostrophic currents from post-processed satellite altimetry. The space-time resolution of these data in the region of interest is much lower than that of COPERNICUS IBI and for this reason we disregarded these choices.

\subsection{The Dynamical Systems Approach}
When a marine oil spill occurs, its fate depends on characteristics such as the oceanic and atmospheric conditions, the quantity of oil released, its physico-chemical properties and other weathering factors. For our analysis we will assume that the spill evolves as if it were composed of passively advected fluid parcels that are not subjected to degrading effects. Such particles follow trajectories $\mathbf{x}(t)$ on the ocean surface that evolve according to the dynamical system:
\begin{equation}
\frac{d\mathbf{x}}{dt} = \mathbf{v}(\mathbf{x}(t),t) \;,
\label{ds_eq}
\end{equation}
where $\mathbf{v}(\mathbf{x}(t),t)$ is the surface ocean velocity field, retrieved from COPERNICUS IBI. This approach is rather reasonable for IFO 380 fuel oil which is a bit denser than crude oil and moves mainly horizontally, close to the surface, a bit below the waterline and thus not subjected to direct wind sailing effects. However, wind effects are considered because the velocity fields $\mathbf{v}(\mathbf{x}(t),t)$ from COPERNICUS IBI take into account atmospheric forcings (see section 2.2 for more details). The accuracy of this approach is supported, as justified in the next section, by the agreement between the predictions made by the simulations, the oil sightings from satellite and in situ observations.

\begin{figure}[htbp]
\centering
\includegraphics[scale=0.6]{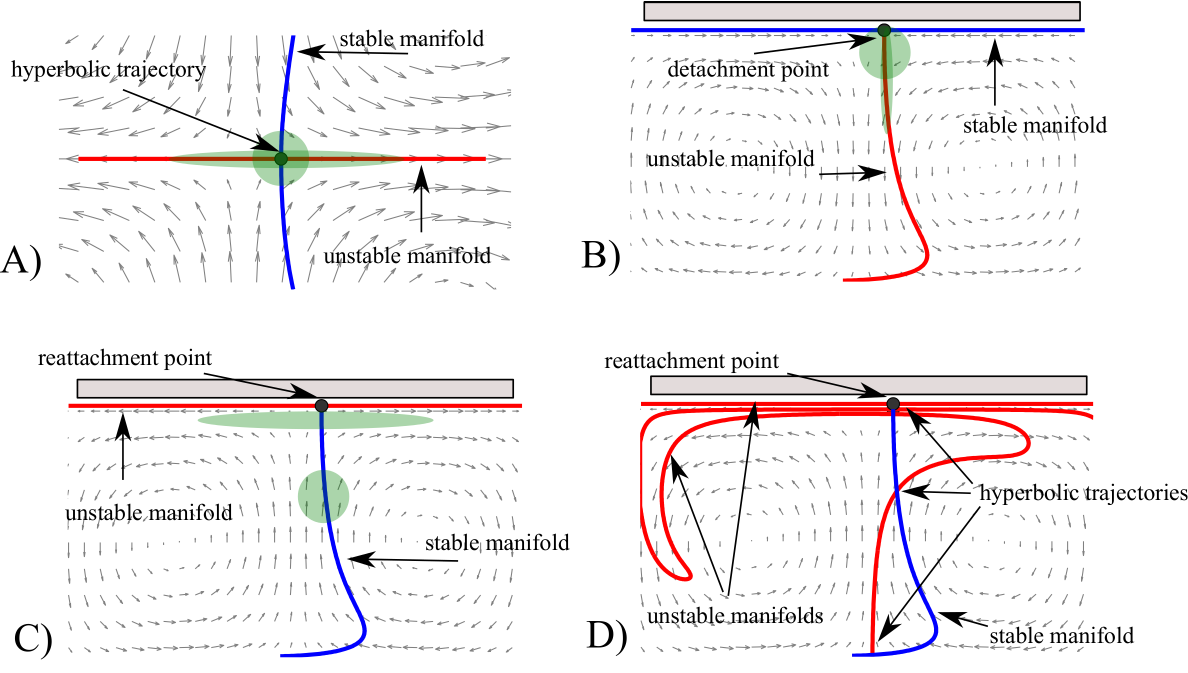}
\caption{Sketches of different hyperbolic trajectories and their stable and unstable manifolds. A) An interior hyperbolic trajectory; B) a detachement point; C) a reattachment point; D) a reattachment point and additional unstable manifolds.}
\label{hyp_desc}
\end{figure}

As explained in the introduction, the global behavior of particle trajectories generated by Eq. (\ref{ds_eq}) can be understood through the spatio-temporal template formed by geometrical structures of the flow that organize trajectories into distinct ocean regions, corresponding to qualitatively different types of trajectories. The boundaries, or barriers, between these regions are time-dependent material surfaces which, mathematically, are invariant manifolds. We explain next in more detail these ideas in our setting. An essential feature in our description are the hyperbolic trajectories. These  are  trajectories in the flow, characterized by high contraction and expansion rates. Figure \ref{hyp_desc}A) illustrates this by plotting two green blobs over a hyperbolic trajectory at successive times. The blobs expand along the unstable manifold (in red) and compress along the stable manifold (in blue). For time dependent flows, hyperbolic trajectories do not correspond to hyperbolic instantaneous stagnation points of the velocity field. Their positions can be rather different, thus being the behaviour of particles in a time interval quite counter-intuitive from what is observed in a frozen time velocity field.

Hyperbolic trajectories can appear not only in the interior of the flow, but also attached to the coastline. In this case,  depending on how the stable and unstable manifolds are placed with respect to the boundary, two configurations are possible. For the configuration illustrated in figure \ref{hyp_desc}B) hyperbolic trajectories are given the special name of detachment point. This configuration is related to the phenomena of flow separation. For the configuration  illustrated in figures \ref{hyp_desc}C) and D), hyperbolic trajectories are referred to as reattachment points.  Figure \ref{hyp_desc}C) shows the evolution of blobs in successive times close to a reattachment point. The initial green circle approaches the coast line following the stable manifold, and elongates itself along the unstable manifold. Figure \ref{hyp_desc}D) provides further details of the underlying skeleton, by plotting two unstable manifolds: the one associated to the reattachment point, and another one associated with a hyperbolic trajectory in the interior of the flow. The latter manifold folds wildly close to the  coast, showing a tangled structure which justifies the filamentation phenomena that would influence  passive scalars  transported by this flow. Additional hyperbolic trajectories are found in the intersection of the stable and unstable manifolds. These trajectories evolve in time by approaching the reattachment point. The configuration portrayed through figures \ref{hyp_desc}C) and D) is of particular interest in our setting.  We will explain in the next section how these features are present in the ocean during the days that followed the sinkage of the Oleg Naydenov and played a key role for describing the oil spill evolution.

\begin{figure}[htbp]
\centering
\includegraphics[scale=0.45]{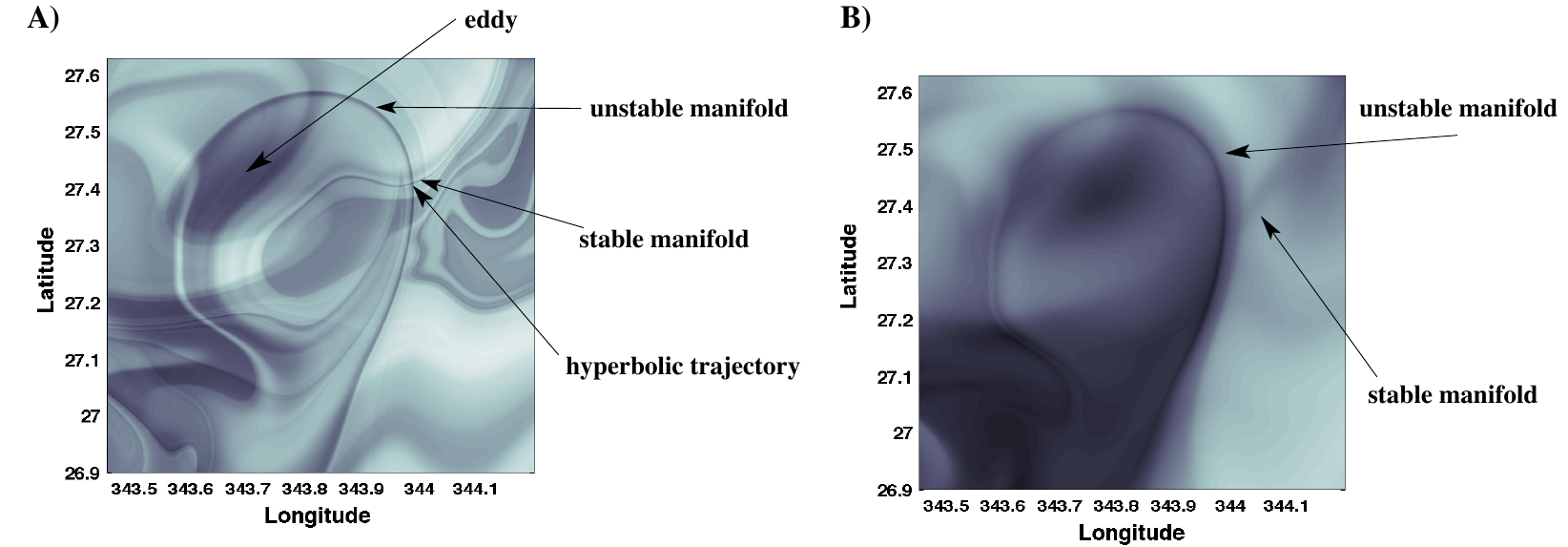}
\caption{Evaluation of the $M$ function close to the Canary Islands using COPERNICUS IBI velocity fields on the 20th April 2015 at 00:00 UTC: A) for $\tau = 15$ days; B) for $\tau = 5$ days. A) and B) highlight the position of visible invariant manifolds and hyperbolic points.}
\label{m_tau}
\end{figure}

A skeleton of the stable and unstable manifolds of hyperbolic trajectories in time dependent flows can be constructed with the technique that is referred to as Lagrangian descriptors, based on the function $M$ (see \cite{mm2009,prl,cnsns}), defined as follows:
\begin{equation}
M(\mathbf{x}_{0},t_0,\tau) = \int_{t_0-\tau}^{t_0+\tau}||\mathbf{v}(\mathbf{x}(t;\mathbf{x}_0),t)|| \; dt \;.
\label{M} 
\end{equation}
Here $||\cdot||$ stands for the modulus of the velocity vector. At a given time $t_0$, the function $M(\mathbf{x}_{0},t_0)$ measures the arc length traced by the trajetory starting at $\mathbf{x}_0 = \mathbf{x}(t_0)$ as it evolves forwards and backwards in time for a time interval $\tau$. Figure \ref{m_tau}A) shows the structure of the function $M$ evaluated for COPERNICUS IBI data close to the Canary Islands on the 20th of April, using an integration period of $\tau = 15$ days.
 Sharp changes of $M$ values (what we call singular features of $M$) occur in narrow gaps, forming filaments that highlight stable and unstable manifolds and, at their crossings, hyperbolic trajectories are visible. Features representing a hyperbolic trajectory and its stable and unstable manifold are marked in the figure. A rule of thumb for making the right interpretation of which lines correspond to stable  and which ones to the unstable manifolds is that stable manifolds never cross stable manifolds (they always appear nested) and similarly occurs with the unstable manifolds. As a result crossing lines  always correspond to  manifolds  with different stability. In this way once a line is identified in the picture as a stable (or unstable) manifold (for instance by using information from the velocity field or from the time evolution) so are all the other lines in the figure. A thorough explanation of how $M$ highlights manifolds is discussed in \citep{prl,mm2012,cnsns,carlos}. Some numerical methods for the direct computation of the manifolds of hyperbolic trajectories in oceanographic contexts and their correspondence with the features revealed by $M$ can be found in \citep{physicad,msw,prl,mm2012}. Figure \ref{m_tau}B) shows the structure of the function $M$ under the same conditions computed for $\tau=5$ days. Now the pattern is smoother, but still some features are perceived. The increasing pattern complexity from B) to A) is justified because $M$ reflects the history of particle trajectories, and in highly chaotic systems this history is expected to be more complex for longer time intervals. 

\begin{figure}[htbp]
\centering
A)\includegraphics[scale=0.45]{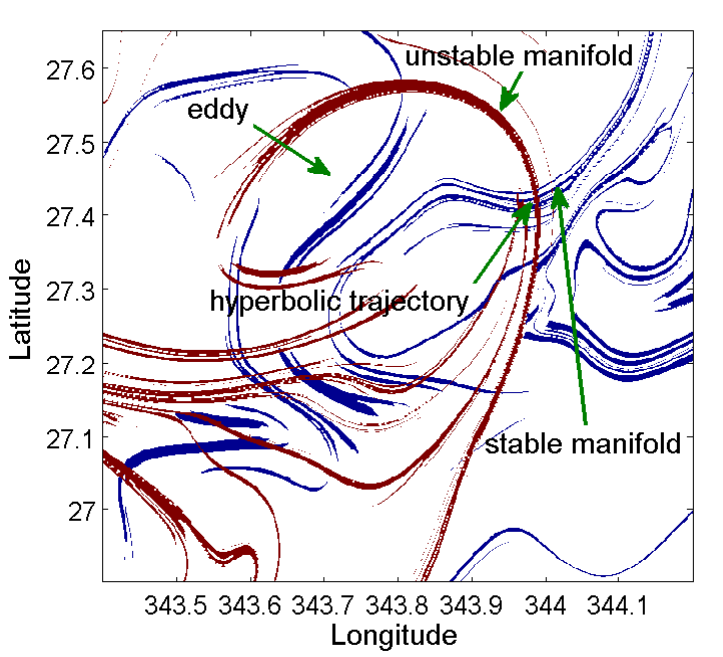}B)\includegraphics[scale=0.45]{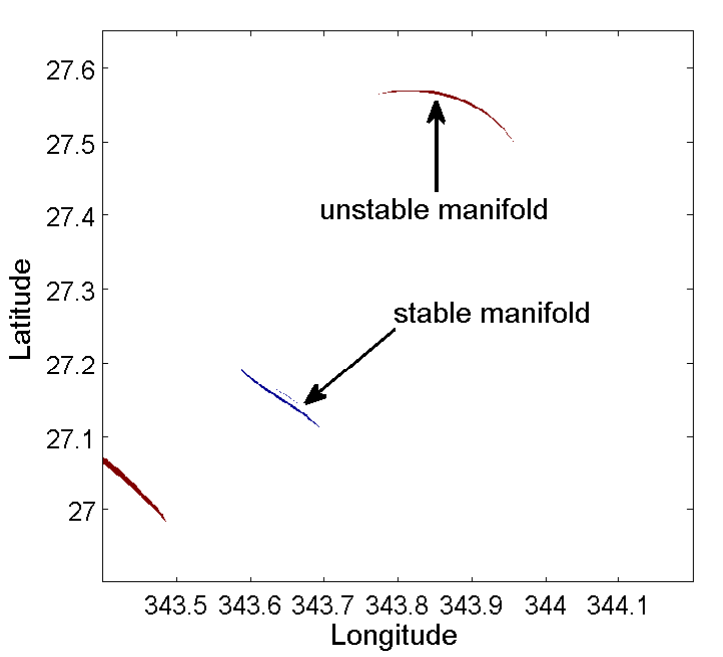}
\caption{Forward (blue) and backward (red)  FTLE fields  obtained for different integration time intervals $\tau$, close to the Canary Islands using COPERNICUS IBI velocity fields on the 20th April 2015 at 00:00 UTC: A)  $\tau = 15$ days and threshold 0.26; B)  $\tau = 5$ days and threshold 0.79. A) and B) highlight the position of visible invariant manifolds and hyperbolic points.}
\label{ftle_tau}
\end{figure}

Another tool for revealing invariant manifolds in time dependent flows, which has been
 extensively applied in geophysical  contexts \cite{olas,Lekien,sha09}, are  FTLE \cite{haller, shaden}.  Shadden et al. \cite{shaden} have proven that fluxes across ridges of FTLE fields are small under certain assumptions, thus confirming their approximate invariant character.  A standard technique to compute FTLE is described, for instance, in \cite{shaden,cnsns}. Forward FTLE, which are calculated  using forward time integration of particle trajectories for a period $\tau$, highlight stable manifolds. Analogously   backward FTLE, obtained from backwards time integration,  highlight unstable manifolds.
 In practice the invariant structures  are obtained from FTLE, not by computing  the ridges of the FTLE fields themselves, but by retaining the values of the FTLE field above a certain threshold (see   \cite{brawigg}). Figure \ref{ftle_tau} displays the forward and backward FTLE fields above a certain threshold for exactly the same cases shown in Figure \ref{m_tau}. As it happens with the $M$ function more details are perceived for larger  integration time intervals $\tau$.  However now, although the same structures are noticed, they are not so clear as those observed from $M$. This effect has been also discussed in \cite{amism11}.

In order to deal with Eq. (\ref{ds_eq}), we have assumed that particle motion is mainly horizontal on a sphere of radius $R$: 
\begin{equation}
\begin{split}
& \dfrac{d\lambda}{dt} = \dfrac{u(\lambda,\phi,t)}{R \cos\phi},\\
& \dfrac{d\phi}{dt} = \dfrac{v(\lambda,\phi,t)}{R},
\end{split}
\label{eqm1}
\end{equation}
where $R = 6371$ km is the Earth's mean radius, $(\lambda,\phi)$ are longitude and latitude, and $(u,v)$ represent, respectively, the zonal (eastward) and meridional (northward) components of the velocity field at the surface, which are specifically the data provided by COPERNICUS IBI. Our Lagrangian analysis is based on the application of LD (i.e. function $M$) and the contour evolution of spills, so we need to integrate (\ref{eqm1}) in order to obtain particle trajectories. Since the velocity field is only given on a discrete space-time grid, the first issue to deal with is that of interpolation. We have used bicubic interpolation in space and third order Lagrange polynomials in time according to the details given in \cite{cf,mmw14}. This assumes smoothness of the velocity field below the resolution. To address the evolution of blobs which could have been potentially related to oil releases, we have tracked the evolution of areas representing spills by means of a contour advection algorithm developed in \cite{dritschel}, but including some modifications explained in \citep{physicad,msw,physrep,malas}.

\section{Results and Discussion}
The sinking point of the fishing ship Oleg Naydenov, highlighted in Fig. \ref{events}B),  was at $15^{\circ}31^{\prime}$ W and $27^{\circ}28^{\prime}$ N. The Advanced Marine Post (AMP) detailed the chronological events that were registered regarding the Oleg Naydenov's spill, between the 23rd April 2015 (the day the spill arrived to the coast of Gran Canaria for the first time) and the 15th May. They also reported the sequence of actions taken by the authorities as a response to the catastrophe (information available at http://www.fomento.gob.es/MFOMBPRENSA/). Figure \ref{events}B) summarizes the confirmed oil sightings, both in the coastline and close to it, as stated by the AMP. The location of the first arrival to the coast on the 23rd April is at the west of the island and it spread towards the north and south during the following days. Dashed lines in Fig. \ref{events}B) outline the daily coast guard aircraft monitoring flight paths from the 23rd April until the 10th May. The green sections correspond to regions where no oil was reported and red color accounts for sections in which oil spill was reported. The blue line in Fig. \ref{events}C) along the Gran Canaria coast represents the coast guard helicopter track on those dates.

\begin{figure}[htbp]
 \centering
 \includegraphics[scale=0.6]{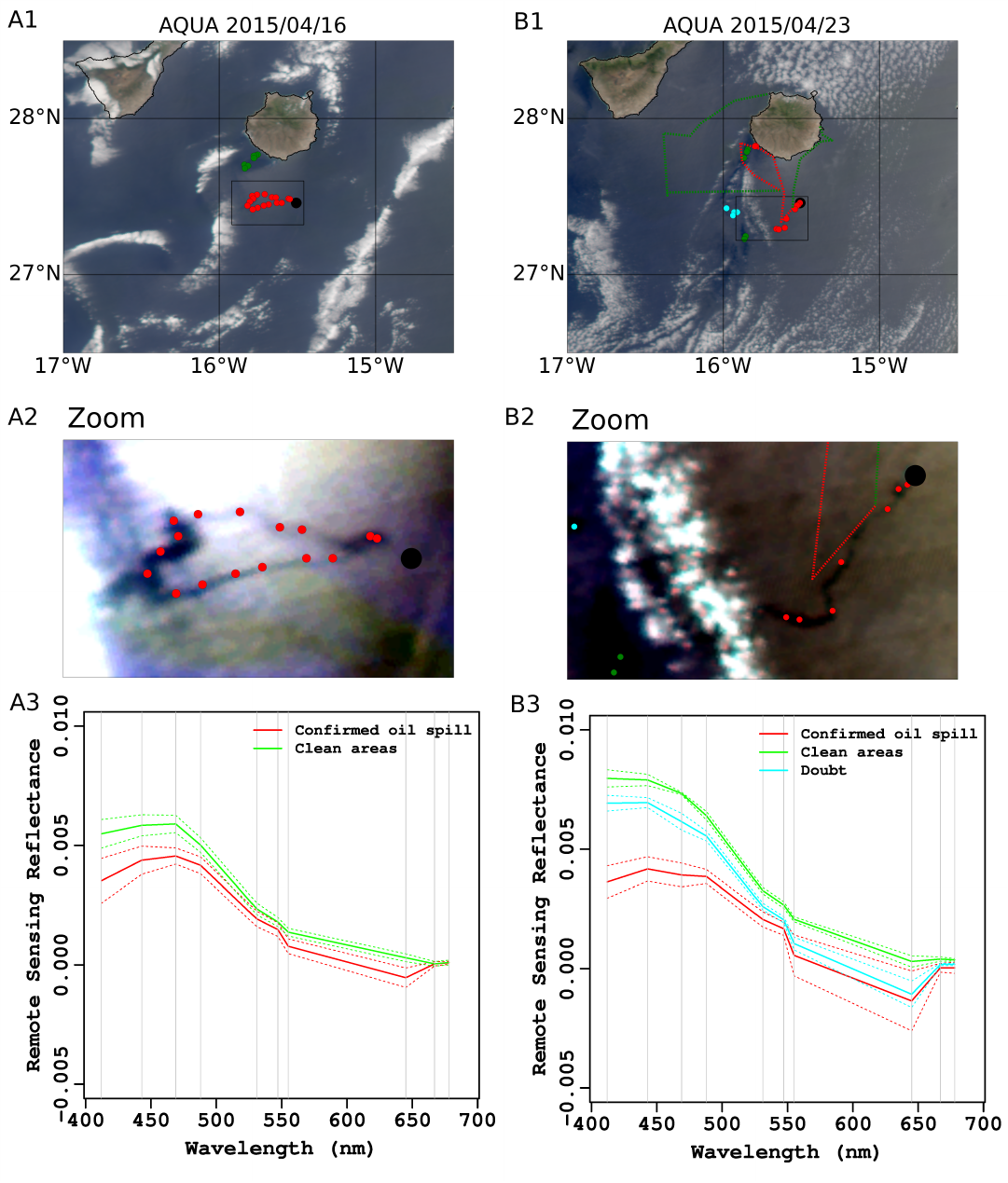}
 \caption{Data derived from MODIS-AQUA corresponding to the 16th April 2015 (column A) and  23rd April 2015 (column B). A1 and B1 account for the quasi-true color images. The sinking point is shown with a filled black circle; A2 and B2 account  for linear enhancement over the zoom of the quasi-true color images. The dashed red and green lines in B1 and B2 account for the daily coast guard aircraft monitoring flight paths (green corresponds to clean ocean areas and red accounts for confirmed oil spill sections); A3 and B3 account for Rrs (Remote Sensing Reflectance) ocean color spectra (mean and standard deviation are displayed as solid and dashed lines respectively), derived from the sampling points in the maps which appear coloured consistently with those shown in the Rrs spectra. The vertical grey lines in A3 and B3 correspond to the MODIS central wavelength bands.}
 \label{montaje}
\end{figure}

The analysis of the satellite images is in agreement with the observations and reports from the aircraft and ground operatives. Our analysis is focused on  MODIS quasi true color images and the processed Remote Sensing Reflectance (Rrs) spectra obtained on the 16th and 23rd April. In the waters surrounding the sinking point, the oil spill is detected in the MODIS quasi-true color images under moderate sunglint conditions (see Fig. \ref{montaje} A1, A2, B1 and B2). However, inside the calm zones of the Gran Canaria island wake (southwest of the island), cloud shadows and aerosol changing patterns make the image interpretation difficult. Moreover, in the case of no sunglint or high sunglint, the visualization of the spill in these kind of images is obstructed. 

The monitoring of the event has been possible, in spite of MODIS limited spatial resolution compared to the width and spread of the oil spill, because the Rrs spectra highlights differences between locations with confirmed oil spills and clean regions (see Fig. \ref{montaje} A3 and B3). Some of the available confirmed points were located near the coast (see Fig. \ref{events}B), where the use of MODIS data for Rrs spectra assessment is less adequate. Bulgarelli and Djavidnia (2012) suggested a particular MODIS processing to derive spectral properties for oil spills in the case of non-glint contaminated areas, and pointed out that the extraction of spectral information over glint contaminated areas is unfeasible. Their work assumed uniform atmospheric properties, specifically, the same aerosol optical thickness. This assumption was not acceptable in the area considered for this study, due to the changing patterns in the atmospheric properties \cite{coca14}. Thus, in order to minimize atmospheric correction problems, only points with moderate sunglint contamination were selected to extract Rrs spectra for Fig. \ref{montaje}. In Fig. \ref{montaje} A2 and B2, the oil spill reported close to the sinking point was clearly visible. However, in the southwestern part of the island, other dark patches appear in the images which were sampled to confirm if oil was present there or not. Rrs spectra were extracted for these patches (green and blue points in Fig. \ref{montaje}) and compared to those corresponding to confirmed oil spill points. The short wavelength spectra (which corresponds to the four lower vertical lines in Fig. \ref{montaje} A3 and B3) showed differences between the oil spill affected areas (Rrs $< 0.005$) and the clean oil ocean regions (Rrs $> 0.005$). In the case of 'doubt' points, MODIS analysis was unable to determine whether they were contaminated or not.

\begin{figure}[htbp]
\centering
\includegraphics[scale=0.5]{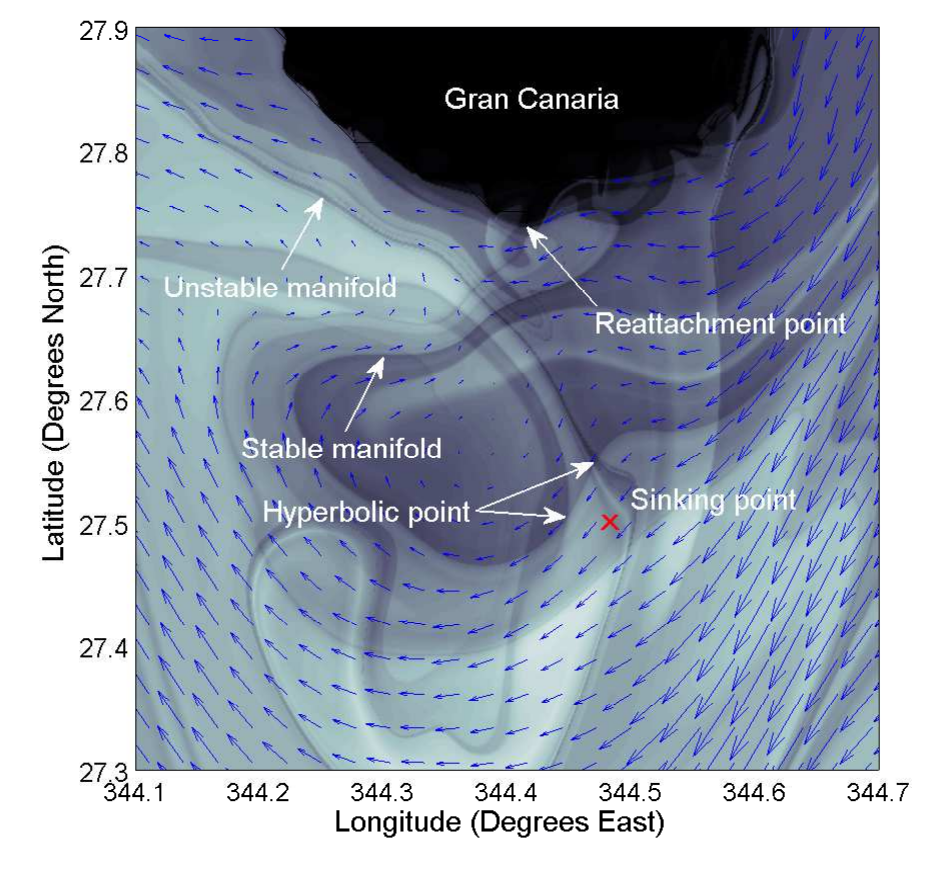}
\caption{Visualization of the function $M$, evaluated for $\tau=15$ days  close to the Gran Canaria coast on the 15th April 2015 at 00:00 UTC. Superimposed, blue arrows show the velocity field at that time, and the sinking point is also marked with a red cross. The reattachment point and its stable manifold are also highlighted. Several hyperbolic points are marked. The wild foldings of the unstable manifolds close to the reattachment point are visible.}
 \label{mcanarias}
\end{figure}

\begin{figure}[htbp]
\centering
\includegraphics[scale=0.5]{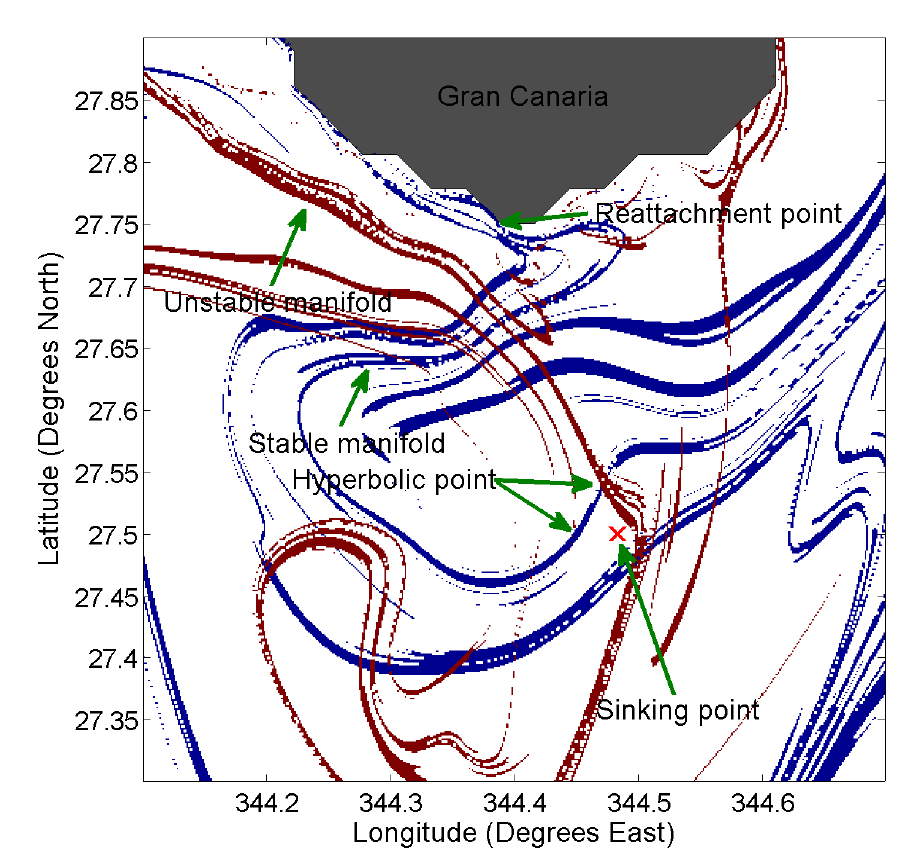}
\caption{Visualization of the forward (blue)  and backward (red) FTLE, evaluated for $\tau=15$ days and threshold 0.26, close to the Gran Canaria coast on the 15th April 2015 at 00:00 UTC. The sinking point is  marked with a red cross. The reattachment point and its stable manifold are also highlighted. Hyperbolic trajectories are distinguishable  at the crossing points. The wild foldings of the unstable manifolds close to the reattachment point are also visible.}
 \label{ftlecanarias}
\end{figure}

The dynamical systems perspective brings new information into the description of these events, providing a dynamical template that helps to understand the long-time behavior of fuel blobs, and thus having the capability to identify potentially dangerous regions for oil spill disasters. In particular we show that the ingredients schematically described in the previous section are also present in the real ocean and provide deep insights into the oil spill evolution. Figure \ref{mcanarias} marks the Oleg Naydenov's sinking point with a cross and also depicts the Lagrangian skeleton governing the dynamical evolution of purely advected particles, as those of fuel-oil, in the ocean region close to the Gran Canaria island on the 15th April. The delicate sinuous lines visible from the contours of the function $M$ highlight the dynamical barriers governing the transport processes in the spirit described in the previous section. In this figure it is highlighted a reattachment point in the coastline of Gran Canaria. This point has a stable manifold which bends towards the sinking point. The stable manifold is crossed by unstable manifolds, forming additional hyperbolic points, unfortunately some of them very close to the Oleg Naydenov's sinking point. This picture confirms that Oleg Naydenov's sinking point was potentially very damaging, as it was close to a dynamical transport highway directly connecting to the coast. Also, Fig. \ref{mcanarias} displays the velocity field of that day superimposed, offering a comparison between the prediction methods used from an Eulerian description and those coming from the Lagrangian approach. Observe how the frozen velocity field fails to reveal the full dynamical structures that underly the flow. A full visualization of the time evolution of these transport barriers is available from movie S1. In the movie it is observed that the reattachment point is a  moving saddle which evolves towards the north of the coast of Gran Canaria. The hyperbolic points which in Figure \ref{mcanarias} are placed close to the sinking point, evolve  in time staying on the stable manifold (as it must be)  approaching the reattachment point, exactly as explained in Fig. \ref{hyp_desc} D).
FTLE fields provide a similar setting, as observed from figure \ref{ftlecanarias}, although as before the Lagrangian features are rough and not so clearly perceived. For this reason  we limit  our further discussions to outputs obtained from function $M$.

The validity of the framework we have just described is confirmed by the sequence of events that occurred after the 15th April 2015. After the sinkage oil slicks were spotted on the sea surface (10-15 liter/minute). At the zero zone the spill is modeled assuming that the fuel released from the ship's tank, in its upwards path towards the surface, spreads over an area which we have fixed to be a circle of 6 km radius. 
The chosen radius  is rather consistent with the perimeter of the spill confirmed on the 16th April (40 hours after the sinkage) which was around 95km, as measured from satellite images (see Fig. \ref{montaje} A2), and that would represent a circle of 15 km radius. Although on the other hand it is slightly bigger than the width of the spill estimated from in situ observations during the first few days of the event, which was around 2 km. Later in this section we will discuss on the impact that the radius choice has on the predictions.

\begin{figure}[htbp]
 \centering
 A)\includegraphics[scale=0.36]{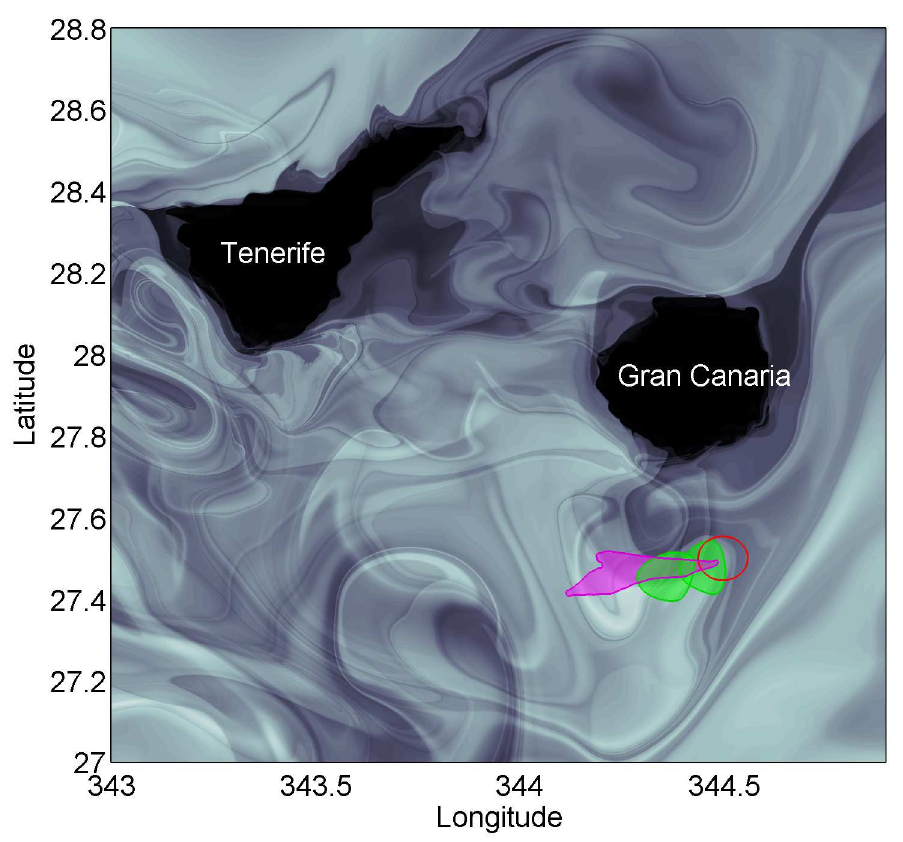}
 B)\includegraphics[scale=0.36]{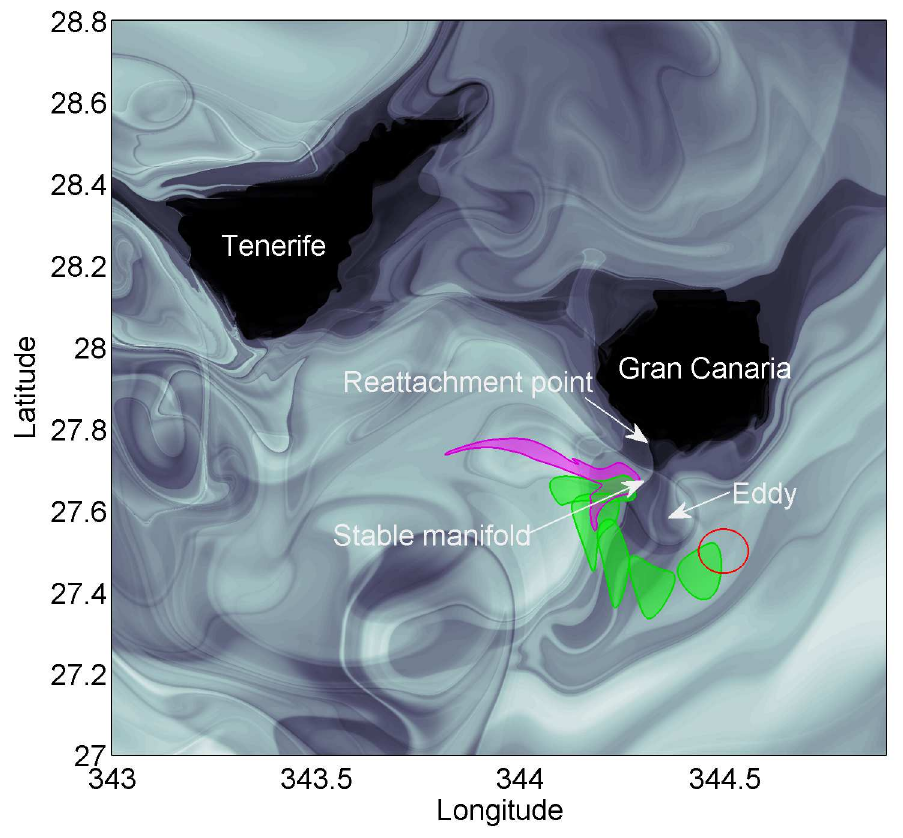}\\
 C)\includegraphics[scale=0.36]{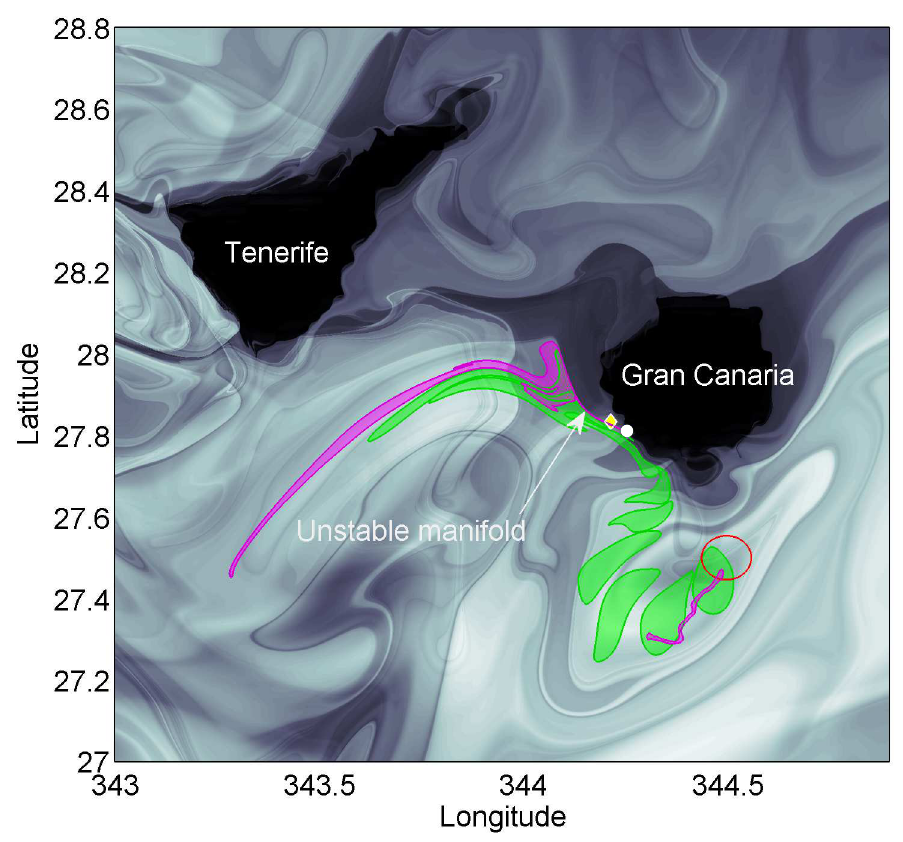}
 D)\includegraphics[scale=0.36]{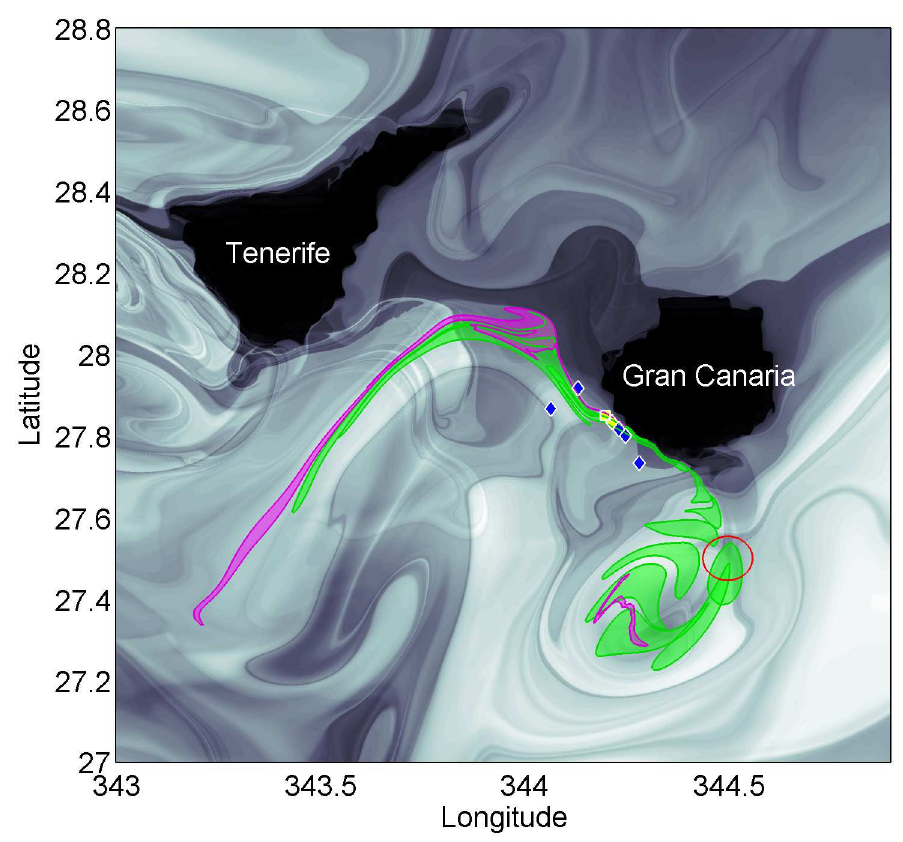}
 \caption{Time evolution of the green (modelled) and pink (confirmed) spills: A) the 16th April; B) the 19th April; C) the 23rd April; D) the 25th April. The yellow diamonds represent fuel sightings on the 23rd April, yellow squares on the  24th April, and blue diamonds on the 25th April. The white circle highlights the position of the reattachment point on the 23rd April.}
 \label{mspill}
\end{figure}

For the analysis of the spill's fate, we have modelled a fuel release every 24 hours, including in this way any potential discharge along the days prior to the sealing of the tank. The contours of these spills are evolved according to the advection equations (\ref{eqm1}) and their time evolution is visible in green color in movie S2. As noticed in Fig. \ref{montaje} A2, on the 16th and 23rd April fuel spills were confirmed from the satellite images, by means of MODIS reflectance spectra (short wavelengths with a Rrs $< 0.005$). These spills have also been used as initial inputs in our simulations. Their time evolution is depicted in pink color in movie S2. The first frame in the movie where the pink contour appears, corresponds to a confirmed observation, but its progress is calculated according to the advection equations (\ref{eqm1}). Several remarks are suggested by this movie. The first one, clear from Fig. \ref{mspill}A, is that both the green and pink spills overlap in a large area on the 16th  April. The second one, distinguishable in Fig. \ref{mspill}B on the 19th April, is that during the first stages, the spill evolves circulating around the outer part of an eddy, pursuing a stable manifold of a reattachment point on the Gran Canaria coast. Spills eventually hit the coast on the 23rd April, spreading along the coast following the foldings of the unstable manifolds as Fig. \ref{mspill}C illustrates. It is remarkable that the arrival day according to the model coincides with the first observed fuel sighting in the coast, marked in Fig. \ref{mspill}C with a yellow diamond. Also it is noteworthy that the  yellow diamond in turn is very close to the position of the reattachment point marked in this figure with a white circle. The spills are strongly compressed across the stable manifold and stretched along the unstable manifold, forming filamentous structures, which on the 25th April 2015 are highly correlated to confirmed fuel sightings. This is appreciable from Fig. \ref{mspill}D. The filamentous structures noticeable in the channel between Gran Canaria and Tenerife were not confirmed in any official report, but the model predicts their presence. Probably, these very thin filamentous structures went unnoticed because they corresponded to a degraded fuel, difficult to detect, or also it could be that the model fails in this aspect. Figure \ref{mspill}C also highlights in pink the spill confirmed in Fig. \ref{montaje}B. The evolution of the spills near the sinking point between the 23rd and the 25th April follow a circulating path trapped in the interior of an eddy  southwest of the sinking point (see Fig. \ref{mspill}D). 

\begin{figure}[htbp]
\centering
\includegraphics[scale=0.5]{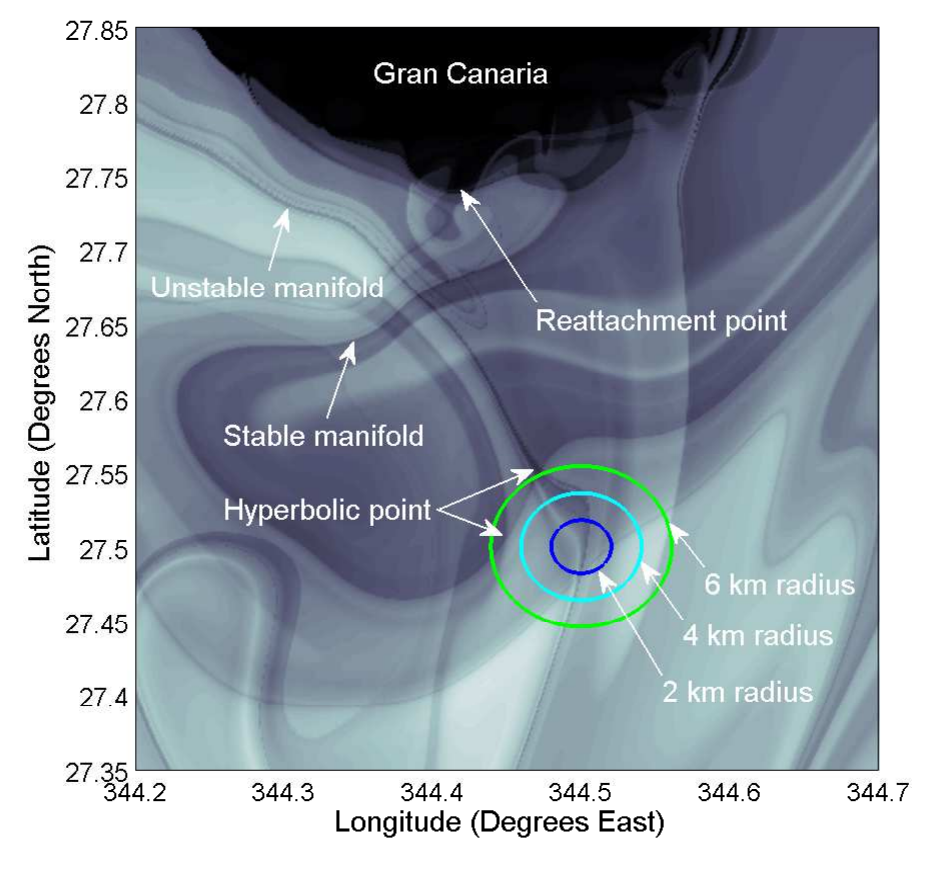}
\caption{Different circles of initial conditions considered for the simulation of the spill. Superimposed, contours of the function $M$, computed for $\tau=15$ days, on the 15th April 2015 at 00:00 UTC. The circles are centered at the point where the oil surfaced after the sinkage. Observe that the proximity of the center of these circles to an hyperbolic point and the stable manifold, that could lead the fuel to the coast via the reattachment point, highlights the sensitivity of the problem to the choice of initial conditions. }
 \label{iconds}
\end{figure}

As remarked before, according to the composition supplied by the stable and unstable manifolds of hyperbolic trajectories, the Oleg Naydenov sinking point was very close to a critical position. Fig. \ref{iconds} shows circles with different radii centred at the sinking point. Observe that the proximity of the center of these circles to two hyperbolic points and the stable manifold that leads the fuel to the coast via the reattachment point, highlights the sensitivity of the problem to the choice of initial conditions. Only the blob of 6 km radius crosses the stable manifold of the reattachment point, and some of the initial conditions within this blob hit the coast on the 23rd April. The 2km and 4km radii choices do not predict any arrival of the oil to the coast. It is evident then that the skeleton provided the $M$ field  allows a visual understanding of the criticality of the setting. On the other hand it should be clear that the important issue here it is not about the exact size of the radius, but the fact that typically 
Lagrangian patterns are robust versus perturbations of the velocity field. This implies that small variations of the velocity field (due to the intrinsic uncertainties in the modelling processes) would lead to similar patterns although maybe shifted or slightly distorted, and then clearly the proximity of the sinking point to these structures, announces a likely fatal outcome.

\begin{figure}[htbp]
\centering
A) \hspace{-.26cm} \includegraphics[scale=0.37]{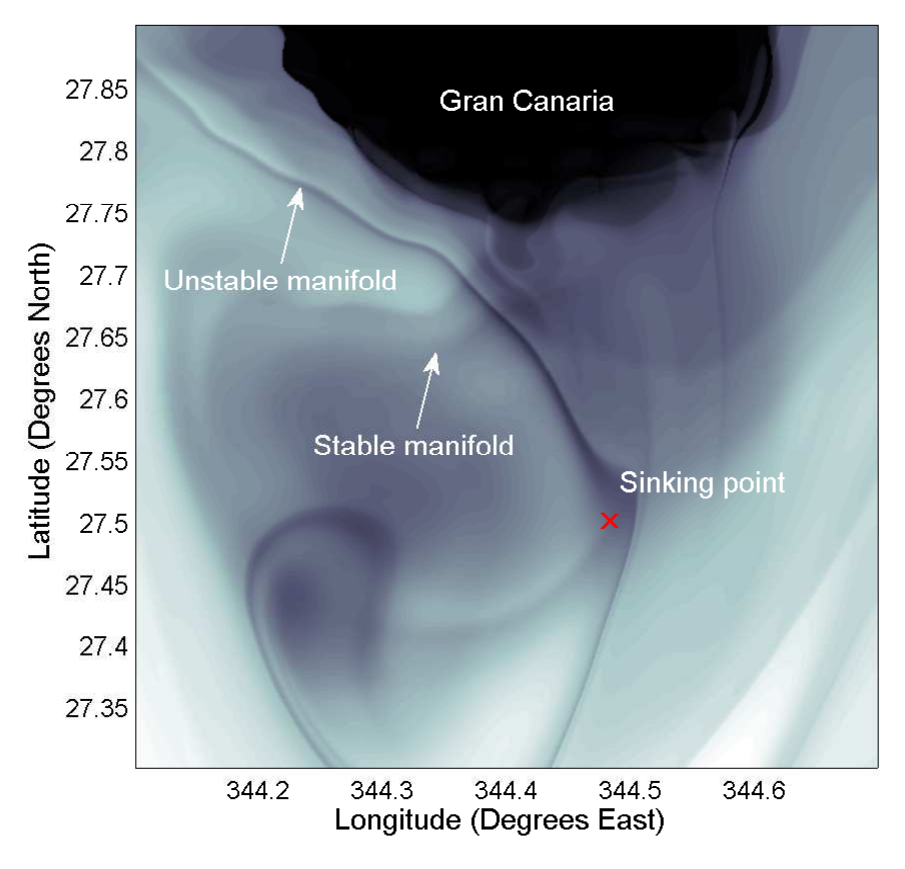} \hspace{-.3cm}
B) \hspace{-.26cm} \includegraphics[scale=0.37]{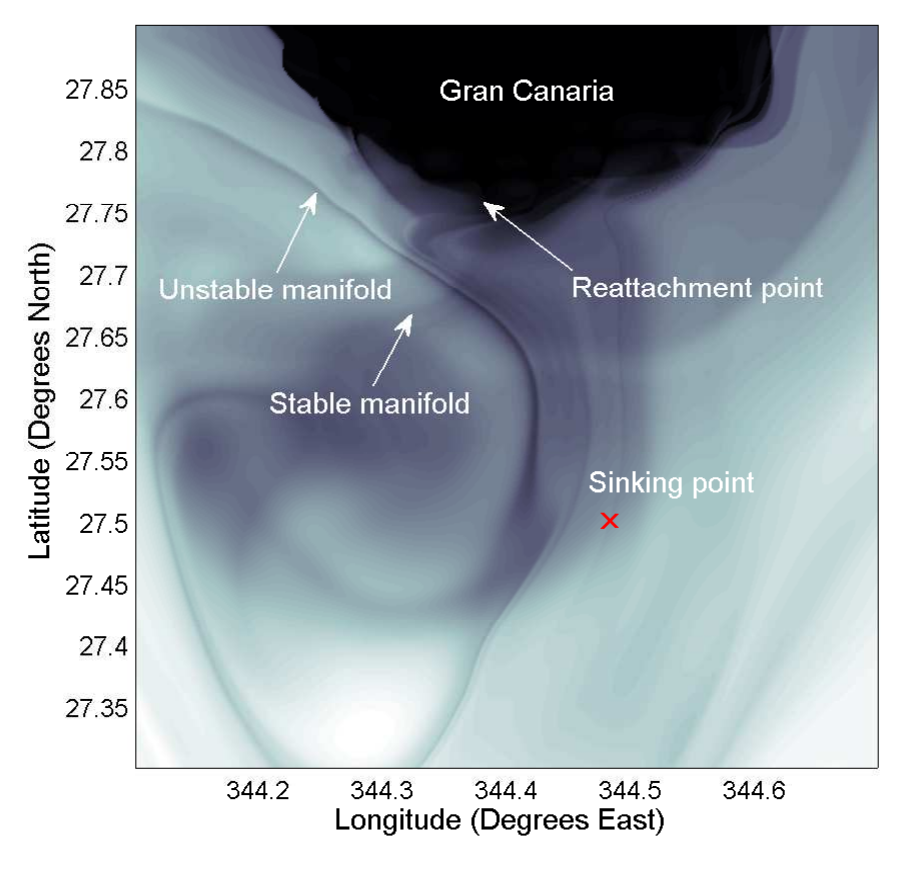}
\caption{Visualization of the function $M$, computed for $\tau=5$ days, close to the Gran Canaria coast: A) the 15th April 2015; B) the 16th April 2015. The proximity of the sinking point to a stable manifold, and the presence of a reattachment point at the coast clearly indicate the highly dangerous nature of the event for the Canary Islands' ecosystem. The highlight of these features for a low $\tau$ value, in this case $\tau = 5$ (five day forecast  required for the calculations), supports the operational character of this methodology.}
 \label{operational}
\end{figure}

For clarity, and to illustrate the complex geometrical patterns revealed in the flow by Lagrangian descriptors, we have computed $M$ in this paper with $\tau = 15$ days as the integration period. However, for operational purposes this choice is not practical as it would require a forecast far beyond the predictions given by the COPERNICUS data set. For an operational approach, a lower time integration value ($\tau = 5$) should be used. This is an accesible value, since  a five day forecast is provided by COPERNICUS IBI. Figure \ref{operational} shows that with this choice, the dynamical features are faded but the dominant ones are still visible. The output pattern highlights the potential danger of the sinking point, and this knowledge could have contributed to the design of decision-making strategies to manage this event.

\section{Conclusions}
In this paper we have brought together tools from different disciplines, such as remote sensing techniques and  dynamical systems theory, to analyse the Oleg Naydenov oil spill that took place in the Canary Islands during April 2015. Our purpose is to demonstrate the effectiveness of these tools for emergency plans and decision-making strategies in order to deal with future marine oil spills in real-time. 

In order to achieve this goal we have described the event from several perpectives. First we have described the resources mobilised by the emergency services  and we have reported a synthesis of the information gathered from ground locations by the numerous observations of the daily aerial and sea monitoring operatives. Second, we have analyzed images of MODIS aboard the Aqua and Terra satellites, which were the most appropriate satellite images for this particular event. MODIS images have been confirmed versus in situ observations, and a procedure to discriminate true from false oil spills using the Rrs spectra has been discussed. Significantly, MODIS images have supplied feedback to the numerical simulations by  providing re-intializing data on two particular days. Finally we have simulated the event under the assumption that the fuel oil is advected as a passive scalar by velocity fields taken from COPERNICUS-IBI. Under this approximation forward blob simulations have been produced. Also geometrical templates are calculated by means of the function $M$. The reliability of the simulations is evidenced by the remarkable agreement between the observations of reported oil sightings and modelled filamentous patches, the fact that the day and place the oil impacted the coast coincide with the observed ones, etc. In particular, these facts confirm that advection accurately represents the evolution of the fuel oil and that the velocity fields taken from COPERNICUS-IBI succesfully represent the true ocean movement, confirming the high quality of this product. As a consequence, these velocity fields are trustworthy and so is any further information based on them such as the one revealed by $M$. This backs the conclusions visualized from $M$ regarding the potential danger of the sinking point. Indeed, the lagrangian skeleton highlights, close to the sinking point, a line which corresponds to a stable manifold of a reattachment point. This means that any blob hitting that line will end up in the coast. 

The agreement of satellite observations and numerical simulations with ground observations confirm that these tools have the potential to be effective contributors to the decision-making process of the emergency services in the real-time management of oil spills. 

\section*{\bf Acknowledgements}
We would like to thank the Regional Committee of Emergencies: Oleg Naydenov 16th April - 30th May 2015 and the Direcci\'on General de Seguridad Mar\'itima (Ministerio de Fomento, Spain) for providing the daily technical reports of the event that were used for this paper. We would also like to thank the OceanColor Web for providing the MODIS data and SeaDAS package (http://oceancolor.gsfc.nasa.gov/) and the R-Cran project (http://www.r-project.org/) for providing the R software. Finally we would also like to thank the European COPERNICUS IBI system for providing access to the database of currents for this study.

V. J. Garc\'ia-Garrido and A. M. Mancho  are supported by MINECO grant MTM2014-56392-R. A. Ramos and J. Coca are supported by MINECO grant UNLP-13-3E-2664 (2013-2015). The research of S. Wiggins is supported by ONR grant No. N00014- 01-1-0769. We acknowledge support from MINECO: ICMAT Severo Ochoa project SEV-2011-0087. Thanks are owed to CESGA and ICMAT for computing facilities.

\bibliography{LD}

\begin{thebibliography}{10}
\expandafter\ifx\csname url\endcsname\relax
  \def\url#1{\texttt{#1}}\fi
\expandafter\ifx\csname urlprefix\endcsname\relax\def\urlprefix{URL }\fi
\expandafter\ifx\csname href\endcsname\relax
  \def\href#1#2{#2} \def\path#1{#1}\fi

\bibitem{pisano}
A.~Pisano, F.~Bignami, R.~Santoleri, Oil spill detection in glint-contaminated
  near-infrared {MODIS} imagery, Remote Sens. 7~(1) (2015) 1112--1134.
\newblock \href {http://dx.doi.org/10.3390/rs70101112}
  {\path{doi:10.3390/rs70101112}}.

\bibitem{bul}
B.~Bulgarelli, S.~Djavidnia, On {MODIS} retrieval of oil spill spectra
  properties in the marine environment, IEEE Geosci. Remote S. 9~(3) (2012)
  398--402.
\newblock \href {http://dx.doi.org/10.1109/LGRS.2011.2169647}
  {\path{doi:10.1109/LGRS.2011.2169647}}.

\bibitem{mp1}
Q.~Xu, X.~Li, Y.~Wei, Z.~Tang, Y.~Cheng, W.~G. Pichel, Satellite observations
  and modeling of oil spill trajectories in the bohai sea, Marine Pollution
  Bulletin 71~(1-2) (2013) 107--116.

\bibitem{mp2}
M.~Marta-Almeida, M.~Ruiz-Villarreal, J.~Pereira, P.~Otero, M.~Cirano,
  X.~Zhang, R.~D. Hetland, Efficient tools for marine operational forecast and
  oil spill tracking, Marine Pollution Bulletin 71~(1-2) (2013) 139--151.

\bibitem{mp3}
Y.~Cheng, X.~Li, Q.~Xu, O.~Garcia-Pineda, O.~B. Andersen, W.~G. Pichel, Sar
  observation and model tracking of an oil spill event in coastal waters,
  Marine Pollution Bulletin 62~(2) (2011) 350--363.

\bibitem{vulpiani}
E.~Aurell, G.~Boffeta, A.~Crisanti, G.~Paladin, A.~Vulpiani, Predictability in
  the large: An extension of the concept of {L}yapunov exponent, J. Phys. A
  :Math. Gen. 30 (1997) 1--26.

\bibitem{nese}
J.~M. Nese, Quantifying local predictability in phase space, Physica D 35~(1)
  (1989) 237--250.
\newblock \href {http://dx.doi.org/10.1016/0167-2789(89)90105-X}
  {\path{doi:10.1016/0167-2789(89)90105-X}}.

\bibitem{shaden}
S.~C. Shadden, F.~Lekien, J.~E. Marsden, Definition and properties of
  {L}agrangian coherent structures from finite-time lyapunov exponents in
  two-dimensional aperiodic flows, Physica D 212~(3-4) (2005) 271--304.
\newblock \href {http://dx.doi.org//j.physd.2005.10.007}
  {\path{doi:/j.physd.2005.10.007}}.

\bibitem{sha09}
S.~C. Shadden, F.~Lekien, J.~D. Paduan, F.~P. Chavez, J.~E. Marsden, The
  correlation between surface drifters and coherent structures based on
  high-frequency radar data in monterey bay, Deep Sea Res. II 56~(3-5) (2009)
  161--172.
\newblock \href {http://dx.doi.org/10.1016/j.dsr2.2008.08.008}
  {\path{doi:10.1016/j.dsr2.2008.08.008}}.

\bibitem{emilio}
F.~d'Ovidio, V.~Fern\'andez, E.~Hern\'andez-Garc\'ia, C.~L\'opez, Mixing
  structures in the {M}editerranean sea from finite--size {L}yapunov exponents,
  Geophys. Res. Lett. 31 (2004) L17203.

\bibitem{birds}
E.~T. Kai, V.~Rossi, J.~Sudre, H.~Weimerskirch, C.~Lopez, E.~Hernandez-Garcia,
  F.~Marsac, V.~Gar{\c c}on, Top marine predators track lagrangian coherent
  structures, Proceedings of the National Academy of Sciences of the USA (PNAS)
  106 (2009) 8245--8250.

\bibitem{olas}
F.~J. Beron-Vera, M.~J. Olascoaga, An assessment of the importance of chaotic
  stirring and turbulent mixing on the west florida shelf., J. Phys. Oceanogr.
  39~(7) (2009) 1743--1755.

\bibitem{physrep}
A.~M. Mancho, D.~Small, S.~Wiggins, A tutorial on dynamical systems concepts
  applied to {L}agrangian transport in oceanic flows defined as finite time
  data sets: {T}heoretical and computational issues, Phys. Rep. 237~(3-4).
\newblock \href {http://dx.doi.org/10.1016/j.physrep.2006.09.005}
  {\path{doi:10.1016/j.physrep.2006.09.005}}.

\bibitem{msw}
A.~M. Mancho, D.~Small, S.~Wiggins, Computation of hyperbolic trajectories and
  their stable and unstable manifolds for oceanographic flows represented as
  data sets, Nonlin. Proc. Geophys. 11 (2004) 17--33.
\newblock \href {http://dx.doi.org/10.5194/npg-11-17-2004}
  {\path{doi:10.5194/npg-11-17-2004}}.

\bibitem{mm2012}
C.~Mendoza, A.~M. Mancho, The lagrangian description of aperiodic flows: {a}
  case study of the {K}uroshio current, Nonlin. Proc. Geophys. 19~(14) (2012)
  449--472.
\newblock \href {http://dx.doi.org/10.5194/npg-19-449-2012}
  {\path{doi:10.5194/npg-19-449-2012}}.

\bibitem{jpo}
A.~M. Mancho, E.~Hern\'andez-Garc\'{\i}a, D.~Small, S.~Wiggins, V.~Fern\'andez,
  {L}agrangian transport through an ocean front in the {N}orth-{W}estern
  {M}editerranean sea, J. Phys. Oceanogr. 38~(6) (2006) 1222--1237.

\bibitem{nlpg2}
C.~Mendoza, A.~M. Mancho, M.-H. Rio, The turnstile mechanism across the
  {K}uroshio current: analysis of dynamics in altimeter velocity fields,
  Nonlin. Proc. Geophys. 17~(2) (2010) 103--111.

\bibitem{beron-vera}
G.~Haller, F.~J. Beron-Vera, Geodesic theory of transport barriers in
  two-dimensional flows, Physica D 241~(7) (2012) 1680--1702.

\bibitem{fh12}
M.~Farazmand, G.~Haller, Computing lagrangian coherent structures from
  variational lcs theory, Chaos 22 (2012) 013128.

\bibitem{rypina}
I.~I. Rypina, S.~E. Scott, L.~J. Pratt, M.~G. Brown, Investigating the
  connection between complexity of isolated trajectories and {L}agrangian
  coherent structures, Nonlin. Proc. Geophys. 18 (2011) 977--987.

\bibitem{mezic3}
I.~Mezic, S.~Wiggins, A method for visualization of invariant sets of dynamical
  systems based on the ergodic partition, Chaos 9~(1) (1999) 213--218.
\newblock \href {http://dx.doi.org/10.1063/1.166399}
  {\path{doi:10.1063/1.166399}}.

\bibitem{mezic}
I.~Mezic, S.~Loire, V.~A. Fonoberov, P.~A. Hogan, A new mixing diagnostic and
  {G}ulf oil spill movement, Science 330~(6003) (2010) 486--489.
\newblock \href {http://dx.doi.org/10.1126/science.1194607}
  {\path{doi:10.1126/science.1194607}}.

\bibitem{gary}
G.~Froyland, K.~Padberg-Gehle., Almost-invariant and finite-time coherent sets:
  directionality, duration, and diffusion., Ergodic Theory, Open Dynamics, and
  Coherent Structures. Proceedings in Mathematics and Statistics 70 (2014)
  171--216.

\bibitem{froy12}
G.~Froyland, C.~Horenkamp, V.~Rossi, N.~Santitissadeekorn, A.~S. Gupta,
  Three-dimensional characterization and tracking of an agulhas ring, Ocean
  Modelling 52-53~(69-75).

\bibitem{Lekien}
F.~Lekien, C.~Coulliette, A.~J. Mariano, E.~H. Ryan, L.~K. Shay, G.~Haller,
  J.~Marsden, Pollution release tied to invariant manifolds: A case study for
  the coast of florida, Physica D 210~(1-2) (2005) 1--20.
\newblock \href {http://dx.doi.org/10.1016/j.physd.2005.06.023}
  {\path{doi:10.1016/j.physd.2005.06.023}}.

\bibitem{prl}
C.~Mendoza, A.~M. Mancho, The hidden geometry of ocean flows, Phys. Rev. Lett.
  105~(3) (2010) 038501.
\newblock \href {http://dx.doi.org/10.1103/PhysRevLett.105.038501}
  {\path{doi:10.1103/PhysRevLett.105.038501}}.

\bibitem{cnsns}
A.~M. Mancho, S.~Wiggins, J.~Curbelo, C.~Mendoza, Lagrangian descriptors: {A}
  method for revealing phase space structures of general time dependent
  dynamical systems, Commun. Nonlinear Sci. 18~(12) (2013) 3530--3557.
\newblock \href {http://dx.doi.org/10.1016/j.cnsns.2013.05.002}
  {\path{doi:10.1016/j.cnsns.2013.05.002}}.

\bibitem{rempel}
E.~L. Rempel, A.~C.-L. Chian, A.~Brandenburg, P.~R. Munuz, S.~C. Shadden,
  Coherent structures and the saturation of a nonlinear dynamo, J. Fluid Mech.
  729 (2013) 309--329.
\newblock \href {http://dx.doi.org/10.1017/jfm.2013.290}
  {\path{doi:10.1017/jfm.2013.290}}.

\bibitem{alvaro}
A.~de~la C\'amara, C.~R. Mechoso, K.~Ide, R.~Walterscheid, G.~Schubert, Polar
  night vortex breakdown and large-scale stirring in the southern stratosphere,
  Clim. Dynam. 35~(6) (2009) 965--975.
\newblock \href {http://dx.doi.org/10.1007/s00382-009-0632-6}
  {\path{doi:10.1007/s00382-009-0632-6}}.

\bibitem{alvaro2}
A.~de~la C\'amara, C.~R. Mechoso, A.~M. Mancho, E.~Serrano, K.~Ide,
  Quasi-horizontal transport within the {A}ntarctic {P}olar-{N}ight vortex:
  {R}ossby wave breaking evidence and lagrangian structures, J. Atmos. Sci. 70
  (2013) 2982--3001.
\newblock \href {http://dx.doi.org/10.1175/JAS-D-12-0274.1}
  {\path{doi:10.1175/JAS-D-12-0274.1}}.

\bibitem{dritschel}
D.~G. Dritschel, Contour dynamics and contour surgery: {N}umerical algorithms
  for extended, high-resolution modelling of vortex dynamics in
  two-dimensional, inviscid, incompressible flows, Comput. Phys. Rep. 10~(3)
  (1989) 77--146.
\newblock \href {http://dx.doi.org/10.1016/0167-7977(89)90004-X}
  {\path{doi:10.1016/0167-7977(89)90004-X}}.

\bibitem{physicad}
A.~M. Mancho, D.~Small, S.~Wiggins, K.~Ide, Computation of stable and unstable
  manifolds of hyperbolic trajectories in two-dimensional, aperiodically
  time-dependent vectors fields, Physica D 182~(3) (2003) 188--222.
\newblock \href {http://dx.doi.org/10.1016/S0167-2789(03)00152-0}
  {\path{doi:10.1016/S0167-2789(03)00152-0}}.

\bibitem{madec}
G.~Madec, NEMO ocean engine, Note du P\^ole de mod\'elisation, Institut
  Pierre-Simon Laplace (IPSL) No. 27, France, 2014, version 3.6.

\bibitem{bell}
M.~J. Bell, R.~M. Forbes, A.~Hines, Assessment of the foam global data
  assimilation system for realtime ocean forecasting, J. Marine Syst. 25~(1)
  (2000) 1--22.
\newblock \href {http://dx.doi.org/10.1016/S0924-7963(00)00005-1}
  {\path{doi:10.1016/S0924-7963(00)00005-1}}.

\bibitem{holt}
J.~T. Holt, I.~D. James, An s coordinate density evolving model of the
  {N}orthwest {E}uropean {C}ontinental {S}helf: 1. {M}odel description and
  density structure., J. Geophys. Res-Oceans 106~(7) (2001) 14015--14034.
\newblock \href {http://dx.doi.org/10.1029/2000JC000304}
  {\path{doi:10.1029/2000JC000304}}.

\bibitem{sot08}
M.~G. Sotillo, E.~A. Fanjul, S.~Castanedo, A.~J. Abascal, J.~Menendez,
  M.~Emelianov, R.~Olivella, E.~Garc\'ia-Ladona, M.~Ruiz-Villarreal, J.~Conde,
  M.~G\'omez, P.~Conde, A.~D. Gutierrez, R.~Medina, Towards an operational
  system for oil-spill forecast over {S}panish waters: {I}nitial developments
  and implementation test, Mar. Pollut. Bull. 56~(4) (2008) 686--703.
\newblock \href {http://dx.doi.org/10.1016/j.marpolbul.2007.12.021}
  {\path{doi:10.1016/j.marpolbul.2007.12.021}}.

\bibitem{sot15}
M.~G. Sotillo, S.~Cailleau, P.~Lorente, B.~Levier, R.~Aznar, G.~Reffray,
  A.~Amo-Baladr{\'o}n, J.~Chanut, B.~Mounir, E.~A. Fanjul, The {M}y{O}cean
  {IBI} {O}cean {F}orecast and {R}eanalysis {S}ystems: {O}perational products
  and roadmap to the future {C}opernicus {S}ervice, J. Oper. Oceanogr. 8~(1)
  (2015) 63--79.
\newblock \href {http://dx.doi.org/10.1080/1755876X.2015.1014663}
  {\path{doi:10.1080/1755876X.2015.1014663}}.

\bibitem{maraldi}
C.~Maraldi, J.~Chanut, B.~Levier, N.~Ayoub, P.~D. Mey, G.~Reffray, F.~Lyard,
  S.~Cailleau, M.~Dr\'evillon, E.~A. Fanjul, M.~G. Sotillo, P.~Marsaleix.,
  {NEMO} on the shelf: {A}ssessment of the {I}beria-{B}iscay-{I}reland
  configuration, Ocean Sci. 9~(4) (2013) 745--771.
\newblock \href {http://dx.doi.org/10.5194/os-9-745-2013}
  {\path{doi:10.5194/os-9-745-2013}}.

\bibitem{lorente}
P.~Lorente, S.~Piedracoba, E.~A. Fanjul, Validation of high-frequency radar
  ocean surface current observations in the {NW} of the {I}berian {P}eninsula,
  Cont. Shelf Res. 92 (2015) 1--15.
\newblock \href {http://dx.doi.org/10.1016/j.csr.2014.11.001}
  {\path{doi:10.1016/j.csr.2014.11.001}}.

\bibitem{mm2009}
J.~A.~J. Madrid, A.~M. Mancho, Distinguished trajectories in time dependent
  vector fields, Chaos 19 (2009) 013111.
\newblock \href {http://dx.doi.org/10.1063/1.3056050}
  {\path{doi:10.1063/1.3056050}}.

\bibitem{carlos}
C.~Lopesino, F.~Balibrea, S.~Wiggins, A.~M. Mancho., Lagrangian descriptors for
  two dimensional, area preserving autonomous and nonautonomous maps., Comm.
  Nonlinear Sci. 27~(1-3) (2015) 40--51.
\newblock \href {http://dx.doi.org/10.1016/j.cnsns.2015.02.022}
  {\path{doi:10.1016/j.cnsns.2015.02.022}}.

\bibitem{haller}
G.~Haller, Finding finite-time invariant manifolds in two-dimensional velocity
  fields, Chaos 10~(1) (2000) 99--108.

\bibitem{brawigg}
M.~Branicki, S.~Wiggins, Finite-time {L}agrangian transport analysis: stable
  and unstable manifolds of hyperbolic trajectories and finite-time {L}yapunov
  exponents, Nonlin. Proc. Geophys. 17 (2010) 1--36.

\bibitem{amism11}
A.~de~la C\'amara, A.~M. Mancho, K.~Ide, E.~Serrano, C.~Mechoso, Routes of
  transport across the {A}ntarctic polar vortex in the southern spring, J.
  Atmos. Sci. 69~(2) (2012) 753--767.

\bibitem{cf}
A.~M. Mancho, D.~Small, S.~Wiggins, A comparison of methods for interpolating
  chaotic flows from discrete velocity data, Comput. Fluids 35~(4) (2006)
  416--428.
\newblock \href {http://dx.doi.org/10.1016/j.compfluid.2005.02.003}
  {\path{doi:10.1016/j.compfluid.2005.02.003}}.

\bibitem{mmw14}
C.~Mendoza, A.~M. Mancho, S.~Wiggins, Lagrangian descriptors and the assesment
  of the predictive capacity of oceanic data sets, Nonlin. Proc. Geophys. 21
  (2014) 677--689.
\newblock \href {http://dx.doi.org/10.5194/npg-21-677-2014}
  {\path{doi:10.5194/npg-21-677-2014}}.

\bibitem{malas}
V.~J. Garc\'ia-Garrido, A.~M. Mancho, S.~Wiggins, C.~Mendoza, A dynamical
  systems approach to the surface search for debris associated with the
  disappearance of flight {MH370}, Nonlin. Proc. Geophys. 22~(6) (2015)
  701--712.
\newblock \href {http://dx.doi.org/10.5194/npg-22-701-2015}
  {\path{doi:10.5194/npg-22-701-2015}}.

\bibitem{coca14}
J.~Coca, T.~Ohde, A.~Redondo, L.~Garc\'ia-Weil, M.~Santana-Casiano,
  M.~Gonz\'alez-D\'avila, J.~Ar\'istegui, E.~F. Nuez, A.~G. Ramos, Remote
  sensing of the {El Hierro} submarine volcanic eruption plume, International
  Journal of Remote Sensing 35~(17) (2014) 6573--6598.
\newblock \href {http://dx.doi.org/10.1080/01431161.2014.960613}
  {\path{doi:10.1080/01431161.2014.960613}}.

\end{thebibliography}

\end{document}